\documentclass[prl,twocolumn,showpacs,superscriptaddress,preprintnumbers]{revtex4}

\usepackage{graphicx}
\usepackage{dcolumn}
\usepackage{bm}

\begin{document}

\preprint{APS/123-QED}

\title{Perovskite Manganites Hosting Versatile Multiferroic Phases 

with Symmetric and Antisymmetric Exchange Strictions
}

\author{S. Ishiwata}
 \email{ishiwata@riken.jp}
 \affiliation{Cross-Correlated Materials Research Group (CMRG), ASI, RIKEN, Wako 351-0198, Japan}

\author{Y. Kaneko}%
\affiliation{Multiferroics Project, ERATO, Japan Science and Technology (JST), c/o RIKEN, Wako 351-0198, Japan}%

\author{Y. Tokunaga}%
\affiliation{Multiferroics Project, ERATO, Japan Science and Technology (JST), c/o RIKEN, Wako 351-0198, Japan}

\author{Y. Taguchi}%
\affiliation{Cross-Correlated Materials Research Group (CMRG), ASI, RIKEN, Wako 351-0198, Japan}%

\author{T. Arima}%
\affiliation{Instisute of Multidisciplinary Research for Advanced Materials, Tohoku University, Sendai 980-8577, Japan}%

\author{Y. Tokura}
\affiliation{Cross-Correlated Materials Research Group (CMRG), ASI, RIKEN, Wako 351-0198, Japan}
\affiliation{Multiferroics Project, ERATO, Japan Science and Technology (JST)}
\affiliation{Department of Applied Physics, University of Tokyo, Hongo, Tokyo 113-8656, Japan}%

\date{\today}

\begin{abstract}
Complete magnetoelectric (ME) phase diagrams of orthorhombic $R$MnO$_{3}$ with and without magnetic moments on the $R$ ions have been established. Three kinds of multiferroic ground states, the $ab$-cycloidal, the $bc$-cycloidal, and the collinear $E$-type phases, have been identified by the distinct ME responses. The electric polarization of the $E$-type phase dominated by the symmetric spin exchange ($\bm{S}_{i} \cdot \bm{S}_{j}$) is more than 10 times as large as that of the $bc$-cycloidal phase dominated by the antisymmetric one ($\bm{S}_{i} \times \bm{S}_{j}$), and the ME response is enhanced near the bicritical phase boundary between these multiferroic phases of different origins. These findings will provide an important clue for the development of the magnetically induced multiferroics. 
\end{abstract}

\maketitle

It has been a long standing problem how to enhance the correlation between magnetism and ferroelectricity in a solid. Possible solutions can be found in recent studies on the magnetically induced multiferroics \cite{Tokura_Science, Cheong}, which can be classified into two types; one is driven by antisymmetric exchange striction in the cycloidal spin structure, typified by TbMnO$_{3}$ \cite{Kimura_TbMnO3, Kenz, Arima_PRL}, and the other is driven by symmetric exchange striction in the commensurate collinear spin structure. Since orthorhombic ($o$-) perovskite $R$MnO$_{3}$ ($R$ $=$ rare earth and Y) contains the both types, it provides an ideal laboratory to compare the respective magnetoelectric (ME) properties and extract the essential ingredients for the development of the magnetically induced multiferroics.

The ferroelectricity with a cycloidal spin order in $o$-$R$MnO$_{3}$ has been found for $R$ $=$ Gd, Tb, Dy, and Eu$_{1-x}$Y$_{x}$ \cite{Kimura_TbMnO3,Goto_PRL,Kimura_PRB2005, Hemberger}. Upon the application of magnetic field $B$ along the $a$ axis in the $ab$-cycloidal phase (in $Pbnm$ notation) of Eu$_{1-x}$Y$_{x}$MnO$_{3}$, the spin-cycloidal plane rotates from $ab$ to $bc$, accompanied by the polarization $P$ rotation from $a$ to $c$ \cite{Yamasaki_EuY,Murakawa}. The origin of the ferroelectricity has been discussed in terms of the spin-current model or the inverse Dzyaloshinskii-Moriya (DM) interaction represented by the relation, $\bm{P} \sim \sum A\bm{e}_{ij}\times (\bm{S}_{i} \times \bm{S}_{j})$ \cite{Katsura, Mostovoy, Dagotto}, in which $\bm{e}_{ij}$ is the unit vector connecting the neighboring spins ($\bm{S}_{i}$ and $\bm{S}_{j}$) and both the spin-orbit and super-exchange interaction are relevant to the coefficient $A$.  On the other hand, $o$-$R$MnO$_{3}$ with $R$ = Ho, Tm, Yb, Lu shows a commensurate collinear spin order with a propagation vector $\bm{q}$ = (0, 1/2, 1), which is so-called $E$-type antiferromagnetic order \cite{Munoz_HoMnO3,Huang_YbMnO3,Okamoto_LuMnO3,Pom_TmMnO3}. This phase is allowed to possess $P$ along $a$ due to symmetric exchange striction, which is independent of the spin-orbit interaction. Sergienko $et$ $al.$ proposed the emergence of substantially large $P$ up to 0.12 C/m$^{2}$ in the $E$-type phase with considering the ferromagnetic nearest-neighbor interaction $J_{1}$ mediated by $e_{g}$ electrons as a major source of the exchange striction \cite{Sergienko, Picozzi}.

\begin{figure}[]
\includegraphics[keepaspectratio,width=5.3 cm]{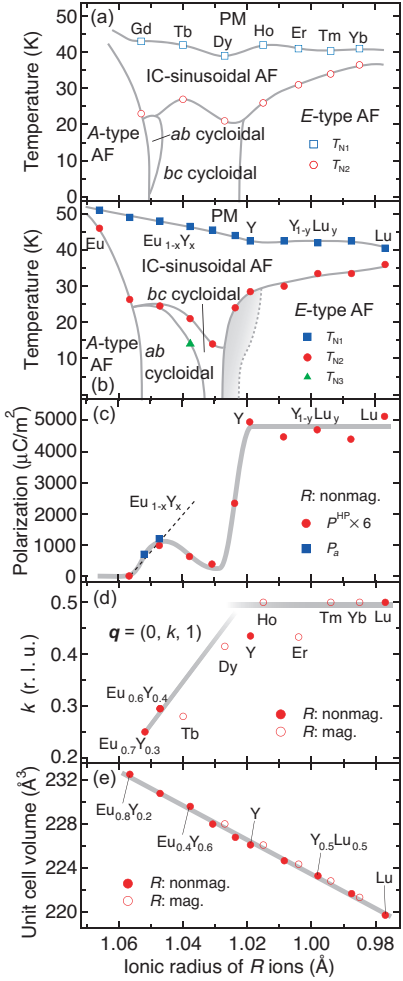}
\caption{\label{fig1} (Color online) Phase diagrams of $o$-$R$MnO$_{3}$ with (a) magnetic $R$ = Gd, Tb, Dy, Ho, Er, Tm, Yb, and (b) nonmagnetic $R$ = Eu$_{1-x}$Y$_{x}$ and Y$_{1-y}$Lu$_{y}$ (PM, IC, and AF denote paramagnetic, incommensurate, and antiferromagnetic, respectively), (c) polarization $P$ value at 2 K for the compounds with nonmagnetic $R$ ions ($P\rm^{HP}$, which is multiplied by a calibration factor (= 6) \cite{footnote}, is for $P$ of the polycrystals, and $P\rm_{a}$ is for $P$ along $a$ of the single crystals taken from ref. \cite{Yamasaki_EuY}, (d) $k$ in the magnetic propagation vector $\bm{q}$ = (0, $k$, 1) (taken from refs. \cite{Arima_PRL,Yamasaki_EuY,Munoz_HoMnO3,Okamoto_LuMnO3,Pom_TmMnO3,Munoz_YMnO3,Ye_ErMnO3,Huang_YbMnO3}), and (e) unit cell volume as a function of the $R$-ion radius, of which coordination number is assumed to be 8 for an orthorhombically distorted perovskite lattice.  $T\rm_{N1}$ was determined by magnetic and dielectric measurements. $T\rm_{N2}$ and $T\rm_{N3}$ were determined from the measurements of $P$. $T\rm_{N1}$ and $T\rm_{N2}$ for $R$MnO$_{3}$ with $R$ = Eu, Gd, and Tb were taken from ref. \cite{Kimura_PRB2005}. The shaded area in (b) represents a possible phase-coexisting region.
}
\end{figure}

The rich variety of magnetic phases of $o$-$R$MnO$_{3}$ reflects the $J_{1}$-$J_{2}$ competition that varies as a function of $R$-ion radius under the staggered orbital ordering of 3$x^{2}-r^{2}$ and 3$y^{2}-r^{2}$ type orbitals; As GdFeO$_{3}$-type lattice distortion becomes large, the antiferromagnetic next-nearest-neighbor interaction $J_{2}$ in the $ab$ plane becomes competitive with the ferromagnetic interaction $J_{1}$. So far, ME phase diagrams of $o$-$R$MnO$_{3}$ have been made for $R$ from La to Dy or Eu$_{0.5}$Y$_{0.5}$ (see Figs. 1(a) and 1(b)) \cite{Kimura_PRB2005, Hemberger,Goto,Tachibana,Mochizuki_RMnO3}. However, since $o$-$R$MnO$_{3}$ with a smaller $R$ ion than Dy needs a high pressure (HP) technique to synthesize, a complete ME phase diagram of $o$-$R$MnO$_{3}$ including the neighboring area between the cycloidal and the $E$-type phases is absent. In fact, although the ferroelectricity in the $E$-type phase has been confirmed for $o$-$R$MnO$_{3}$ with $R$ = Y, Ho \cite{Lorenz2007}, and Tm \cite{Pom_TmMnO3}, intrinsic ME properties inherent to the Mn-spin arrangement alone remain unclear because of the intervention by the magnetic $R$ ions. Besides, the reported $P$ values vary widely depending on the materials. As a result, quantitative estimations of $P$ of the $E$-type phase and its microscopic origin are under intensive debate. 

This Letter reports complete ME phase diagrams of a series of $o$-$R$MnO$_{3}$ with nonmagnetic $R$ ions ($R$ = Eu$_{1-x}$Y$_{x}$ and Y$_{1-y}$Lu$_{y}$) together with the system containing the magnetic $R$ ion ($R$ = Dy, Ho, Er, Tm, Yb). By using high-quality polycrystalline samples of $o$-$R$MnO$_{3}$, we have confirmed substantially large $P$ of nearly 5000 $\mu$C$/$m$^{2}$ for the $E$-type phases, yet one order of magnitude smaller than the predicted values \cite{Sergienko, Picozzi}, and clearly demonstrated the generic transition of the multiferroic ground state from the $ab$-cycloidal, to $bc$-cycloidal, and eventually to the $E$-type phase upon decreasing the size of the $R$ ion. Furthermore, we found an enhanced ME response characteristic of a bicritical phase boundary formed by the different multiferroic phases.

Polycrystalline samples of $o$-$R$MnO$_{3}$ with $R$ = Dy and $R$ not larger than Y were synthesized under HP by using hexagonal ($h$-) $R$MnO$_{3}$ as precursors. First, single crystals of $h$-$R$MnO$_{3}$ were prepared by a floating-zone method, and then heat-treated for 1 hr in the range of 1323 $\sim$ 1373 K under a HP of 5.5 GPa. Just for comparison between the single crystalline and polycrystalline specimens, polycrystalline samples of $o$-Eu$_{1-x}$Y$_{x}$MnO$_{3}$ with $x$ = 0.2, 0.4, 0.6 were synthesized by heat treatment under HP using polycrystalline precursors prepared by grinding the single crystals of the $o$-phase. The grain-boundary effects on ME measurements are expected to be minimized owing to the high-quality precursors. However, as for $o$-Eu$_{1-x}$Y$_{x}$MnO$_{3}$ with $x$ = 0.75, 0.9, the attempts to grow crystals gave the mixtures of the $h$- and the $o$-phases, which were adopted as the precursors. The linear change in the unit cell volume as a function of the $R$-ion radius ensures the successful syntheses of a series of $o$-$R$MnO$_{3}$ (Fig. 1(e)). For measurements of $P$, gold electrodes were deposited on the polished faces (4 mm$^{2}$) of the platelet samples with typically 0.2 mm in thickness. As a poling procedure, an electric field of 800 V$/$mm was applied at 40 K, followed by cooling to 2 K. The displacement current was measured with increasing temperature at a rate of 5 K$/$min or sweeping $B$ at a rate of 100 Oe/sec, and was integrated as a function of time to obtain $P$. Magnetic susceptibility was measured by a SQUID magnetometer.

Figure 2 shows temperature dependence of $P$ of $o$-$R$MnO$_{3}$ prepared by the HP technique, with highlighting different responses of $P$ to external $B$ for $o$-Eu$_{0.1}$Y$_{0.9}$MnO$_{3}$ and $o$-LuMnO$_{3}$. As exemplified in Figs. 2(a) and 2(b), the compounds undergo a transition to an incommensurate (IC) sinusoidal phase at $T\rm_{N1}$ where $\chi$ takes a maximum. Then, they show a second transition to the $E$-type phase at $T\rm_{N2}$ where $P$ sets in. In Figs. 2(a) and 2(b), $P$ in zero magnetic field is plotted against temperature for nonmagnetic and magnetic $R$ ions, respectively. The compounds with $R$ = Eu$_{0.1}$Y$_{0.9}$, Y$_{1-y}$Lu$_{y}$, Ho, Er, Tm, Yb are supposed to possess the $E$-type phase, and all the compounds but for $R$ = Er and Eu$_{0.1}$Y$_{0.9}$ exhibit fairly large $P$ of about 800 $\mu$C$/$m$^{2}$ at the lowest temperature. By contrast, $R$ = Eu$_{1-x}$Y$_{x}$ ($x$ = 0.4, 0.6, 0.75) and Dy sample possess relatively small $P$ values, implying the different mechanism of $P$ generation. We should note here that $o$-YMnO$_{3}$ and $o$-ErMnO$_{3}$ show the stepwise temperature dependence of $P$. The neutron diffraction studies have suggested the IC $\bm{q}$ vectors of (0, 0.435, 1) and (0, 0.433, 1) for the magnetic ground states in $o$-YMnO$_{3}$ and $o$-ErMnO$_{3}$, respectively \cite{Munoz_YMnO3,Ye_ErMnO3}. Thus, these stepwise changes may indicate the presence of a small amount of the cycloidal or the sinusoidal phase which competes and thus coexists with the $E$-type phase. The signature for such a phase coexistence is explicitly found for $o$-Eu$_{0.1}$Y$_{0.9}$MnO$_{3}$ showing smaller $P$ and the larger effect of applied $B$ on $P$ in comparison with LuMnO$_{3}$ (see Figs. 2(c) and 2(d)). The possible phase coexistence in $o$-YMnO$_{3}$ and $o$-Eu$_{0.1}$Y$_{0.9}$MnO$_{3}$ can be attributed to proximity to the first order phase boundary, as discussed later. As for $o$-ErMnO$_{3}$, a long range order of the $E$-type phase is likely to be amenable to the large magnetic moments of Er ions with planar anisotropy, leading to the stepwise change in $P$.

\begin{figure} []
\includegraphics[keepaspectratio,width=8.4 cm]{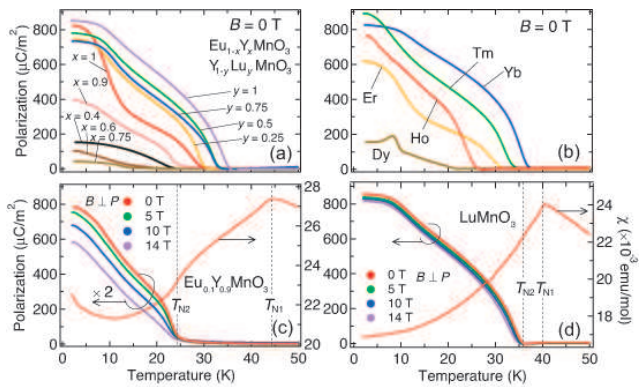}
\caption{\label{fig2} (Color online) Temperature dependence of polarization $P$ of $o$-$R$MnO$_{3}$ with (a) nonmagnetic $R$ (Eu$_{1-x}$Y$_{x}$, Y$_{1-y}$Lu$_{y}$) and (b) magnetic $R$ (Dy, Ho, Er, Tm, and Yb) in the absence of external magnetic field $B$, and (c) $P$ of $o$-Eu$_{0.1}$Y$_{0.9}$MnO$_{3}$ (multiplied by 2) and (d) $P$ of $o$-LuMnO$_{3}$ measured in external $B$ perpendicular to the applied electric field and $\chi$ measured in external $B$ of 0.1 T on increasing temperature. $B$ was applied during the poling procedure as well as the measuring process.
}
\end{figure}

Based on the polarization and magnetization measurements, we established the ME phase diagrams for a series of $o$-$R$MnO$_{3}$ with magnetic and nonmagnetic $R$ ions, respectively (Figs. 1(a) and 1(b)). Here, one may find that the variation of $T\rm_{N1}$ as a function of the ionic radius of $R$ is well systematic when the $R$ ions are nonmagnetic, whereas it is rather irregular when the $R$ ions are magnetic. This difference suggests that the presence of the magnetic moment of the $R$ ions gives substantial influence on the transition temperatures as well as the relative stability of the two kinds of cycloidal phases and the $E$-type phase. Figure 1(c) shows a plot of $P$ at 2 K multiplied by a correction factor (= 6) for the compounds with nonmagnetic $R$ ions \cite{footnote}, together with $P_{a}$ in Eu$_{1-x}$Y$_{x}$MnO$_{3}$, as a function of the $R$-ion radius. Contour plot of the calibrated value of $P$ is shown in Fig. 3(c). In the $E$-type phase, $P$ reaches nearly 5000 $\mu$C$/$m$^{2}$, which is more than 10 times as large as $P$ (= 384 $\mu$C$/$m$^{2}$) in Eu$_{0.25}$Y$_{0.75}$MnO$_{3}$, and almost independent of the $R$-ion radius except for Eu$_{0.1}$Y$_{0.9}$MnO$_{3}$. In the cycloidal phases, on the other hand, $P$ is critically dependent on the $R$-ion radius. According to the inverse DM model, $P$ is proportional to sin$\theta$, where $\theta$ denotes the angle between the nearest neighbor Mn spins. When the $ab$-cycloidal spins are of ideal cycloid and $k$ is less than 0.5 ($\theta$ $<$ $\pi$/2), $P$ ($\propto$ sin$\theta$) is expected to increase monotonically, as indicated by the broken line in Fig. 1(c). However, in the range of 0.4 $<$ $x$ $<$ 0.75, $P$ decreases as increasing $x$. This is probably because the ellipticity of the cycloidal spins becomes larger \cite{Mochizuki_RMnO3}, or a fraction of the $bc$-cycloidal phase with smaller $P$ becomes larger. By further increasing $x$, $P$ shows a local minimum at $x$ = 0.75, where the $bc$-cycloidal state is the magnetic ground state, followed by an abrupt increase that indicates the emergence of the $E$-type phase. 

In order to reveal the difference in ME response of each phase of $o$-$R$MnO$_{3}$ with nonmagnetic $R$, we have investigated $B$ dependence of $P$ at representative points ($T$ = 2 K) marked with open circles in Fig. 3(a). In Eu$_{0.1}$Y$_{0.9}$MnO$_{3}$ and LuMnO$_{3}$ (Figs. 3(d) and 3(e)), the normalized $P$ shows nearly quadratic behavior as a function of $B$ and little dependence on the $B$ direction with respect to the $P$ direction. The results are in accord with the $E$-type phase as the magnetic ground state for both compounds, because $P$ of this phase should be proportional to $\bm{S}_{i} \cdot \bm{S}_{j}$ and thus to 1-$aB^{2}$ when the magnetization is proportional to applied $B$. As clearly seen in Fig. 3(a), the $R$-ion dependence of the transition temperature $T\rm_{N2}$, below which the system becomes ferroelectric, makes a V-shaped feature near the boundary between the $bc$-cycloidal and the $E$-type phases. This feature reminds us of the bicriticality of two competing phases separated by the first-order phase boundary in the presence of weak randomness \cite{Tokura_bicritical}. The $bc$-cycloidal and the $E$-type phases are likely separated by the first-order phase boundary \cite{Furukawa}, and Eu$_{1-x}$Y$_{x}$MnO$_{3}$ has also weak randomness introduced via alloying the $A$ site. Since Eu$_{0.1}$Y$_{0.9}$MnO$_{3}$ is located on the verge of the transition from the $E$-type phase to the $bc$-cycloidal phase,  the system should be subject to the phase coexistence, leading to the enhancement of the effect of $B$ on $P$ (Fig. 3(d)) and the suppression of $P$ (Fig. 2(c)). As a hallmark of the phase coexistence, the large hysteresis was observed in the $P$-$B$ curve of Eu$_{0.1}$Y$_{0.9}$MnO$_{3}$, suggesting that the decrease of $P$ is not only due to the cant of spins in the $E$-type phase but partly due to the increase in the volume fraction of the cycloidal phase which is dominated by the Zeeman energy gain ($\propto \chi B^{2}$). In accord with this expectation, $\chi$ is larger for Eu$_{0.1}$Y$_{0.9}$MnO$_{3}$ containing a small amount of the large-$\chi$ cycloidal phase than for LuMnO$_{3}$ with the purely $E$-type order (see Figs. 2(c) and 2(d)). 

\begin{figure}[]
\includegraphics[keepaspectratio,width= 7.5 cm]{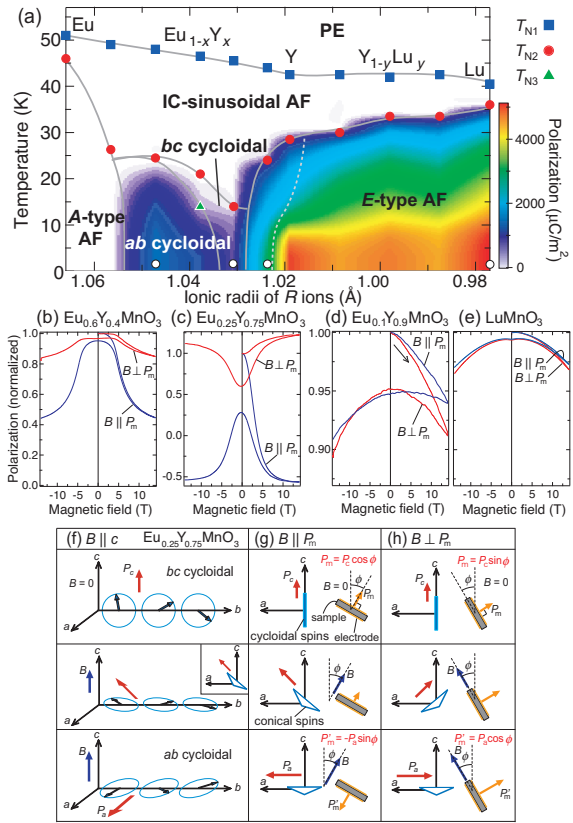}
\caption{\label{fig3} (Color online) (a) Corrected magnitude of polarization $P$ displayed as a contour-plot in the phase diagram of $o$-$R$MnO$_{3}$ with nonmagnetic $R$ (Eu$_{1-x}$Y$_{x}$, Y$_{1-y}$Lu$_{y}$). Magnetic-field $B$ dependence of normalized $P$ at 2 K for (b) Eu$_{0.6}$Y$_{0.4}$MnO$_{3}$, (c) Eu$_{0.25}$Y$_{0.75}$MnO$_{3}$, (d) Eu$_{0.1}$Y$_{0.9}$MnO$_{3}$, and (e) LuMnO$_{3}$ ($P$ was normalized by the value at 2 K). Schematic illustrations of the spin rotation from the $bc$ cycloidal to the $ab$ cycloidal (f) under $\bm{B}$ $\|$ $\bm{c}$, (g) under $\bm{B}$ $\|$ $\bm{P}_{m}$ deviating from $c$ to $a$ by $\phi$, and (h) under $\bm{B}$ $\bot$ $\bm{P}_{m}$ deviating from $c$ to $a$ by $\phi$ ($\phi$ $<$ $\pi$/4). These illustrations depict the spin rotation in a single domain both for lattice and spin. The relational expressions between $P_{m}$ or $P'_{m}$ and $\phi$ are shown.
}
\end{figure}

On the other hand, in the $ab$-cycloidal phase of Eu$_{0.6}$Y$_{0.4}$MnO$_{3}$ (Fig. 3(b)) and the $bc$-cycloidal phase of Eu$_{0.25}$Y$_{0.75}$MnO$_{3}$ (Fig. 3(c)), $P$ changes drastically at critical $B$ of about 5 T and 3 T, respectively, which signifies the occurrence of the $P$ rotation. In fact, the study of a single crystal of Eu$_{0.6}$Y$_{0.4}$MnO$_{3}$ showed the $B$-induced $P$ rotation from $P_{a}$ to $P_{c}$ at 4.5 T ($\bm{B}$ $\|$ $\bm{a}$) \cite{Yamasaki_EuY}. However, because our samples are polycrystals, we have to consider the relative orientations between $B$, $P$, and the crystallographic axes of a single domain. To explain the drastic change in $P$ under $B$, we considered specific situations which satisfy the conditions for the occurrence of the $B$-induced $P$ rotation (hereafter we formally use $P_{m}$ and $P'_{m}$ for measured $P$ before and after the $P$ rotation, respectively). The panels in Fig. 3(f) describe the $P$ rotation from $P_{c}$ to $P_{a}$ upon the application of $B$ along $c$. After the $P$ rotation, two kinds of domains with positive and negative $P_{a}$ coexist in an equal amount (only for the $P$ rotation to positive $P_{a}$ is shown in Fig. 3(f)), so that $P'_{m}$ should be zero. When $B$ is slightly deviated from $c$ to $a$ by $\phi$, finite $P'_{m}$ remains after the $P$ rotation as shown in Figs. 3(g) and 3(h), because the plane of spin spiral inclines to be perpendicular to $B$ to gain the Zeeman energy during the $P$ rotation (see the middle panels), and thereby one of the rotation directions is selected, as reported in Eu$_{0.55}$Y$_{0.45}$MnO$_{3}$ \cite{Murakawa}. If $B$ is slightly deviated from $c$ to $b$, $P'_{m}$ should be zero similarly to the case in Fig. 3(f).  Here, the normalized $P$ after the $P$ rotation ($P'_{m}/P_{m}$) for $\bm{B}$ $\|$ $\bm{P}_{m}$ and $\bm{B}$ $\bot$ $\bm{P}_{m}$ can be expressed as $P'_{m}/P_{m}$ = -tan$\phi$$P_{a}/P_{c}$ and $P'_{m}/P_{m}$ = tan($\pi/2-\phi$)$P_{a}/P_{c}$, respectively. Taking the inequations $P_{a}$ $>$ $P_{c}$ and $\phi$ $<$ $\pi$/4 into account, the drastic decrease (and even sign reversal) of $P$ for $\bm{B}$ $\|$ $\bm{P}_{m}$ and the increase of $P$ for $\bm{B}$ $\bot$ $\bm{P}_{m}$ shown in Fig. 3(c) can be well explained. The same explanation is applicable for the $P$ rotation from $P_{a}$ to $P_{c}$ (Fig. 3(b)).

In summary, we have revealed the magnetoelectric properties of a whole series of multiferroic $o$-$R$MnO$_{3}$ with both nonmagnetic and magnetic $R$ and estimated the genuine values of $P$ ($\sim$ 5000 $\mu$C$/$m$^{2}$) in the $E$-type phase, which is more than 10 times as large as that of the $bc$-cycloidal phase, yet one order of magnitude smaller than the predicted values \cite{Sergienko, Picozzi}. Furthermore, we found the bicritical feature near the phase boundary between the competing $bc$-cycloidal and $E$-type phases. 

\begin{acknowledgments}
The authors thank Y. Takahashi, M. Mochizuki, and N. Furukawa for useful discussions. This work was supported in part by Grants-in-Aid for young scientists (B) (No. 21750069) and Grant-in-Aid for Scientific Research on Priority Areas "Novel States of Matter Induced by Frustration" (No. 20046017)  from the MEXT, Japan.
\end{acknowledgments}


\begin{thebibliography}{28}
\bibitem{footnote}
From the ratio between the polycrystalline and the single crystalline $P$ values in Eu$_{0.6}$Y$_{0.4}$MnO$_{3}$, the correction coefficient to convert the observed $P$ value into the expected value for the single crystal was estimated to be 6. All the $P$ values in Fig. 1(c) and Fig. 3(a) are corrected with this coefficient. This estimation is consistent with the naive expectation that the polycrystalline $P$ value, even if fully poled, should be 1/3 of the single crystalline $P$ value and even smaller since the poling for the polycrystals cannot be complete because of the random poling directions.
\end{thebibliography}
\end{document}